\documentclass[fleqn,usenatbib]{mnras}

\usepackage{newtxtext,newtxmath}
\usepackage[T1]{fontenc}
\usepackage{soul}

\DeclareRobustCommand{\VAN}[3]{#2}
\let\VANthebibliography\thebibliography
\def\thebibliography{\DeclareRobustCommand{\VAN}[3]{##3}\VANthebibliography}

\usepackage{graphicx}	% Including figure files
\graphicspath{Figures/}
\usepackage{tabularx}
\usepackage{booktabs}
\usepackage{amsmath}	% Advanced maths commands
\newcommand{\hi}{H\,{\sc{i}} } %HI
\newcommand{\mhi}{H\sc{i}} % HI for use in maths mode; no space between H and I and no trailing space
\newcommand{\himf}{H{\sc{i}}MF } %HIMF
\newcommand{\himfe}{H{\sc{i}}MF} %HIMF at the end of a sentence

\newcommand{\sqdeg}{deg$^{2}$ }
\newcommand{\sqdegns}{deg$^{2}$}
\newcommand{\kmps}{\text{km\,s}^{-1}}

\newcommand{\ud}{\mathrm{d}} % Command \ud creates upright "d" used in integrals and differentials

\title[Statistical selection of lensed \textsc{Hi} galaxies]{Statistical selection of high-redshift, neutral-hydrogen-rich, lensed galaxies with the Square Kilometre Array}

\author[C. B. Button and R.P. Deane]{
Charissa B. Button,$^{1}$
Roger P. Deane,$^{2,1}$\thanks{E-mail: roger.deane@wits.ac.za}
\\
$^{1}$Department of Physics, University of Pretoria, Lynwood Rd, Pretoria 0028, South Africa\\
$^{2}$Wits Centre for Astrophysics, University of the Witwatersrand, 1 Jan Smuts Avenue, Johannesburg, 2000, South Africa\\
}

\date{Accepted 2025 February 10. Received 2025 February 07; in original form 2024 November 22}

\pubyear{2025}

\begin{document}
\label{firstpage}
\pagerange{\pageref{firstpage}--\pageref{lastpage}}
\maketitle

\begin{abstract}
Deep wide spectral line surveys with the Square Kilometre Array (SKA) will expand the cosmic frontiers of neutral atomic hydrogen (\textsc{Hi}) in galaxies. However, at cosmologically significant redshifts ($z \gtrsim 0.5$), detections will typically be spatially unresolved and limited to the highest mass systems. Gravitational lensing could potentially alleviate these limitations, enabling lower mass systems to be studied at higher redshift and spatially resolved dynamical studies of some \hi discs. Additionally, lensed \hi systems would select foreground dark matter haloes using a different, more extended baryonic tracer compared to other lens surveys. This may result in a wider selected range of foreground dark matter halo properties, such as the concentration parameter. This paper uses the distortion of the observed \hi mass function (\himfe) produced by strong gravitational lensing to find a flux density criterion for selecting lensed \hi sources in future SKA-Mid spectral line surveys. This selection approach could yield lensed \hi source densities in the range of $\sim 0.1$--$10$ galaxies per square degree out to a redshift of $z \simeq 3$ covered by SKA-MID Band 1. Although the sample sizes are modest, even with the proposed SKA-Mid surveys, the selection approach is straightforward and should have a 50\% efficiency without any additional information, such as low-impact-factor or lower-redshift massive galaxies. The efficiency of selecting high-redshift, neutral-hydrogen-rich, lensed galaxies should then be greatly enhanced by using SKA-MID data in concert with the Vera C. Rubin Large Survey of Space and Time.
\end{abstract}


%%%%%%%%%%%%%%%%%%%%%%%%%%%%%%%%%%%%%%%%%%%%%%%%%%

%%%%%%%%%%%%%%%%% BODY OF PAPER %%%%%%%%%%%%%%%%%%

%----------------------------------------------------------------------------------------------
\section{Introduction}
\label{sect: introduction}
%----------------------------------------------------------------------------------------------
Atomic and molecular hydrogen are two fundamental transitory phases in the baryon cycle, which refers to the cycling of gas through different phases (e.g. atomic hydrogen, molecular hydrogen, stars and ionized gas) and through the different regions in and surrounding a galaxy \citep[see e.g.][]{bouche_2010_ImpactCold,lilly_2013_GasRegulation, peroux_2020_CosmicBaryon, walter_2020_EvolutionBaryons}. Understanding the atomic and molecular content within the interstellar medium (ISM) can help to explain global scaling relations such as the star-forming main sequence \citep[][and references therein]{saintonge_2022_ColdInterstellar} and contribute to our theory of star formation and galaxy evolution. Recently, several large surveys have enabled the study of the molecular gas content in galaxies out to high redshifts \citep[e.g.][]{saintonge_2017_XCOLDGASS, tacconi_2018_PHIBSSUnified, decarli_2019_ALMASpectroscopic, decarli_2020_ALMASpectroscopic, magnelli_2020_ALMASpectroscopic}. These surveys include the ALMA Spectroscopic Survey in the Hubble Ultra-deep Field \citep[ASPECS;][]{walter_2016_ALMASpectroscopic} which observed carbon monoxide (CO) as a tracer for $\text{H}_2$ over a large redshift coverage ($z \lesssim 10$), as well as surveys undertaken with the \textit{Herschel Space Observatory}, covering the redshift range $0 \lesssim z \lesssim 2.5$ \citep[see][and references therein]{berta_2016_MeasuresGalaxy}. However, there are comparatively few studies of \hi in similar detail, particularly at higher redshifts, where observations of \hi rely on indirect tracers.

At low redshift, \hi can be observed through its emission line at 21\,cm which arises due to the hyperfine structure levels of hydrogen in its ground state. However, due to the low transition probability, this emission line is weak, thereby limiting our ability to make direct directions at large cosmological distances $(z \gtrsim 0.2)$. Furthermore, because of its fairly long wavelength, the \hi 21-cm line in galaxies beyond the local universe can only be spatially resolved with interferometers.

Over the past two decades there have been several systematic searches for \hi 21-cm emission from low redshift galaxies. Two such surveys, covering wide areas and undertaken with single dish instruments are the \hi Parkes All-Sky Survey \citep[HIPASS;][]{barnes_2001_ParkesAll, meyer_2004_HIPASSCatalogue} and the Arecibo Legacy Fast ALFA survey \citep{giovanelli_2005_AreciboLegacy, haynes_2018_AreciboLegacy}. These two surveys covered redshifts out to $z \lesssim 0.04$ and $z \lesssim 0.06$, respectively. In addition to these single-dish surveys, there have been several surveys undertaken with radio interferometers, including deep \hi surveys that probed higher redshifts than previous studies. The \hi Nearby Galaxies Survey \citep[THINGS;][]{walter_2008_THINGSNearby} studied nearby galaxies using the Very Large Array (VLA) and was able to resolve the \hi content within these galaxies on sub-kpc scales. At higher redshifts, requiring substantial integration times, the Blind Ultra-Deep \hi Environmental Survey \citep[BUDHIES;][]{verheijen_2007_WSRTUltradeep, gogate_2020_BUDHIESIV}, undertaken with the Westerbork Synthesis Radio Telescope (WSRT), and the COSMOS \hi Large Extragalactic Survey \citep[CHILES;][]{fernandez_2013_PilotVery}, undertaken with the VLA, were able to reach the sensitivities required to observe the \hi content in galaxies at redshifts of $z \simeq 0.2$. Recently, the FAST Ultra-Deep Survey \citep{xi_2024_MostDistant}, carried out on the FAST telescope, discovered six \hi galaxies at redshifts between $z \simeq 0.38$ and $z \simeq 0.40$. These examples illustrate the development and improvement in these systematic \hi searches over the past two decades, especially in the aim to detect \hi 21-cm emission at cosmologically significant distances.

At intermediate to high redshifts ($z\gtrsim 0.4$), current radio telescopes are not sensitive enough to detect \hi directly through its 21-cm emission line within $\lesssim24$\,hr (beyond this integration time, systematic calibration errors begin to limit the theoretical sensitivity limits). Hence, high redshift studies rely on indirect methods, such as spectral line stacking, damped Lyman-$\alpha$ absorption studies or \hi absorption studies. Over redshifts of $0.3 \lesssim z \lesssim 1.3$, stacking the \hi 21-cm emission from individual galaxies that are below the detection limit can yield the average \hi content of the sample \citep[e.g.][]{bera_2018_ProbingStar, chowdhury_2020_21centimetreEmission, chowdhury_2021_GiantMetrewave}. However, uncertainties in the redshifts of the galaxies that are used in the stacking, as well as the spatial and spectral apertures over which the individual spectra are extracted, can lead to significant uncertainties in the stacked spectrum and the inferred \hi mass \citep[e.g.][]{elson_2019_UncertaintiesResults}. At higher redshifts ($z \gtrsim 1.6$), \hi can be studied using the Lyman-$\alpha$ transition, which falls into the optical wavebands at these redshifts. Using this approach, \hi is usually studied through absorption in rest-frame UV spectroscopy against a class of quasars known as damped Lyman-$\alpha$ systems (DLAs). This method, however, suffers from line saturation towards high \hi column density systems, limiting its application to the circumgalactic medium \citep{jorgenson_2006_UCSDRadioselected, krogager_2019_EffectDust}. Another approach to studying high-redshift \hi is to study \hi absorption, although this method also suffers from uncertainties due to, for example, spin temperature and gas morphologies. Given the uncertainties in these indirect methods, direct detections of the \hi 21\,cm line at these redshifts are desirable in and of themselves, but could also be used to test and calibrate these indirect approaches.

Due to the natural amplification caused by gravitational lensing, another technique for studying the \hi content in intermediate to high redshift galaxies is to measure \hi 21-cm emission in systems that are gravitationally lensed. Gravitational lensing has played an important role in studying molecular and ionized ISM gas in distant galaxies \citep{solomon_2005_MolecularGas, vieira_2013_DustyStarburst} and could, similarly, offer a promising tool for directly observing the \hi 21-cm line at higher redshift and lower \hi masses than would be possible without lensing \citep{deane_2015_StronglyLensed}. Such observations could enable the \hi scaling relations that are known in the local Universe to be observed and tested as a function of both redshift and \hi mass. They would also enable important cross-checks of the indirect \hi detection methods. Additionally, if the lensed \hi galaxy is resolved, the increase in the angular size will enable resolved studies of the \hi kinematics that would otherwise not be possible, which can be used to extend the current studies of the baryonic Tully--Fisher relation and dark matter distributions at earlier cosmological epochs.

Two of the current MeerKAT \hi surveys, MIGHTEE \citep{jarvis_2016_MeerKATInternational} and LADUMA \citep{blyth_2016_LADUMALooking}, as well as the proposed SKA \hi surveys will detect a large number of \hi sources, on the order of $10^{3}$--$10^{5}$ \citep{staveley-smith_2015_HIScience}. These surveys are also expected to detect lensed systems; for example, LADUMA is predicted to contain around 20 lensed \hi galaxies, while the SKA Deep Survey is expected to contain of order $10^3$ lensed \hi galaxies \citep{deane_2015_StronglyLensed}. Given the much larger number of the unlensed \hi sources in these surveys and the arcsec-scale angular resolution, it remains a challenge to find an efficient method for selecting the lensed sources in the survey data, particularly for lower magnification sources.

Efficiently identifying lens systems in survey data is challenging at all wavelengths. One of the most productive lens searches in optical wavelengths is the SLoan Lens ACS Survey \citep[SLACS;][]{bolton_2008_SloanLens}. The SLACS sample consists of 131 systems where high redshift spectral lines from a distant galaxy are superimposed onto the continuum emission from a lower redshift galaxy along the same line of sight. These spectra were selected from the parent SDSS sample which contained $\sim10^6$ spectra. Follow-up observations with the \textit{Hubble Space Telescope} confirmed lensed nature of at least 70 out of the 131 candidates \citep{bolton_2008_SloanLens}. Another example of a lens search in optical data is the search for strong gravitational lenses in the Dark Energy Spectroscopic Instrument (DESI) Legacy Imaging Surveys Data Release 9 \citep{storfer_2024_NewStrong}. This survey covers a total area of $19\,000\,\text{deg}^2$ and has detected over $\sim 4.5 \times 10^6$ objects. Out of this, 1895 candidate lens systems were identified using a machine learning approach \citep{storfer_2024_NewStrong}. Clearly lens systems are rare, and in this paper we investigate an approach to selecting gravitationally lensed galaxies based only on their \hi 21-cm emission.

Gravitational lensing boosts the observed number counts of a source population, an effect which can be used to select gravitationally lensed objects in large surveys based on a flux density selection criteria. This effect was used to select lensed sub-millimetre galaxies in some of the far infrared and sub-millimetre surveys carried out by \textit{Herschel} \citep{negrello_2017_HerschelATLASSample} and the South Pole Telescope \citep{vieira_2013_DustyStarburst}. For example, \citet{negrello_2017_HerschelATLASSample} extract a sample of 80 lensed sub-mm galaxy candidates that were selected from the \textit{H}-ATLAS survey using an observed frame 500\,$\mu$m flux density threshold of 100\,mJy. This population of sources observed at 500\,$\mu$m is made up of low-redshift late-type galaxies, AGN-powered radio sources and high-redshift sub-mm galaxies. The flux density selection applied to the \textit{H}-ATLAS sample removed the unlensed sub-mm galaxies, while the late-type galaxies and AGN-powered radio sources remained as contaminants and had to be removed using multi-wavelength data. This two-step selection process left a sample that only contained gravitationally lensed sub-mm galaxies and follow-up observations of these lens candidates with either the Submillimeter Array (SMA) or the Atacama Large Millimeter/submillimeter Array (ALMA) confirmed their lensed nature \citep{negrello_2017_HerschelATLASSample}.

In the same way, gravitational lensing will distort the observed \hi mass function (\himfe) due to the amplification of the \hi signal in a subset of sources. Using this effect on the observed \himfe, a flux density threshold can be identified at which the integrated number counts of the lensed sources exceed the integrated number counts of the unlensed sources. This flux density threshold could then, in principle, be used to efficiently select lensed \hi galaxies in the large spectral line surveys of the SKA-Mid. In a previous paper, we have investigated to what extent this same approach would be useful for selecting gravitationally lensed OH megamasers in these same spectral line surveys \citep{button_2024_EfficientSelection}, but the \hi application is significantly different for a multitude of reasons discussed in Section~\ref{sect: distortion himf} and \ref{sect:integrated counts of hi sources}. 

In this paper, this statistical approach of selecting lensed galaxy candidates is applied to the problem of selecting candidate lensed \hi sources in the large upcoming \hi surveys that have been proposed for the SKA-Mid. Specifically, this paper addresses the question as to what extent is this approach practically useful for selecting candidate lensed \hi galaxies in large \hi surveys. Section~\ref{sect: distortion himf} briefly describes how the distortion to the \himf is calculated. Section~\ref{sect:integrated counts of hi sources} discusses the details of how the \hi integrated source counts are calculated. The results of the statistical selection as applied to \hi sources are presented and discussed in Section~\ref{sect: results and discussion}. Finally, the conclusions are presented in Section~\ref{sect: conclusion}. In this paper, we assume a Planck 2018 cosmological model \citep{planckcollaboration_2020_Planck2018} unless otherwise stated.

%----------------------------------------------------------------------------------------------
\section{Distortion of the \himfe}
\label{sect: distortion himf}
%----------------------------------------------------------------------------------------------
The distortion of a mass function or luminosity function due to the effects of gravitational lensing is described in detail in \citet{button_2024_EfficientSelection}, which is a spectral line application of the far infrared continuum approach developed by \citet{perrotta_2002_GravitationalLensing}, \citet{negrello_2007_AstrophysicalCosmological}, and \citet{wardlow_2013_HerMESCandidate}. For the sake of clarity, the essential equations are repeated here. 

The probability that a background source at redshift, $z_{\text{S}}$ is lensed by a magnification factor greater than $\mu$ is given by the ratio of the total lens cross-section area, $\sigma$, to the area of a sphere centred on the earth with radius equal to the angular diameter distance to the source. Since the total lens cross-section is the sum of the lens cross-sections due to all the intervening lenses, this probability can be written as
\begin{equation}
\begin{split}
    P(>\mu, z_{\text{S}}) = & \frac{1}{4 \pi D^{2}_{\text{A}}(z_{\text{S}})} 
    \int_0^{4 \pi} \ud \Omega  \int_0^{z_{\text{S}}} \ud z_{\text{L}} \frac{\ud V_{\text{C}}}{\ud \Omega \ud z_{\text{L}}}\\ 
    & \int \ud M_{\text{vir}} \, \sigma(\mu, z_{\text{L}}, z_{\text{S}}, M_{\text{vir}}) N_{\text{C}}(M_{\text{vir}, z_{\text{L}}}),
    \label{eqn: prob source is lensed vol int}
\end{split}
\end{equation}
where $D_{\text{A}}(z_{\text{S}})$ is the angular diameter distance at the redshift of the source, $\ud V_{\text{C}}/\ud \Omega \ud z$ is the comoving volume element, $\sigma(\mu, z_{\text{L}}, z_{\text{S}}, M_{\text{vir}})$ is the lens cross-section for a single lens system, and $N_{\text{C}}(M_{\text{vir}, z_{\text{L}}})$ is the comoving number density of the lenses. Consistent with previous work \citep[e.g.][]{perrotta_2002_GravitationalLensing, wardlow_2013_HerMESCandidate}, we assume that the lens population consists of dark matter haloes with singular isothermal sphere (SIS) density profiles and we ignore any effect of substructure. Hence, the lens cross-section only depends on the nature of the lens via its virial mass and the comoving number density of the lenses can be be described by the Sheth and Tormen dark matter halo mass function \citep{sheth_2001_EllipsoidalCollapse}.

Given the probability, $P(\mu, z_{\text{S}})$, that a background source is magnified by a factor greater than $\mu$, the magnification probability distribution is given by
\begin{equation}
    p(\mu, z_{\text{S}}) = -\frac{\ud P(\mu, z_{\text{S}})}{\ud \mu}.
    \label{eqn: prob distribution}
\end{equation}

The distorted \himf is then given by
\begin{equation}
    \Phi'(M_{\text{\mhi}}, z_{\text{S}}) = \int_{\mu_{\text{min}}}^{\mu_{\text{max}}} \ud \mu p(\mu, z_{\text{S}}) \Phi \left(\frac{M_{\text{\mhi}}}{\mu}, z_{\text{S}} \right),
    \label{eqn:HIMF distoted by mag bias}
\end{equation}
where $\Phi$ is the intrinsic \himf expressed as the number density of \hi sources per base-10 logarithmic \hi mass interval, and $\Phi'$ is the distorted \himf due to the effect of gravitational lensing on the observed number counts. Here, the minimum magnification, $\mu_{\text{min}}$, is assumed to have a value of $\mu_{\text{min}}=2$ as per the formal definition of strong lensing, while the maximum magnification, $\mu_{\text{max}}$, depends on the angular extent of the background source and is typically estimated using ray-tracing simulations.

Throughout this work, we assume that the unlensed \himf is given by a Schechter function where the parameters are estimated from the ALFALFA sample \citep{jones_2018_ALFALFAHI} and adjusted to the Planck 2018 cosmological parameters \citep{planckcollaboration_2020_Planck2018}. These adjusted parameters are $\alpha = -1.25$, $\log M_{\text{\mhi}}^* = 9.96$ and $\phi^* = (4.24) \times 10^{-3}\, \text{Mpc}^{-3} \, \text{dex}^{-1}$. The largest uncertainties in the parameter values determined by \citet{jones_2018_ALFALFAHI} are in the values of $\alpha$ and $\phi^*$ where the uncertainties are 9\% and 22\%, respectively, while the uncertainty in the characteristic mass, $M_{\text{\mhi}}^*$, is only 0.6\%. The two parameters that most affect the results of this work are the characteristic mass, $M_{\text{\mhi}}^*$, and the normalization, $\phi^*$. Additionally, uncertainties in the results of this paper, arise from the assumed \hi mass--peak flux density model and the assumption of the SIS density profile for the lenses.

Furthermore, in this work we assume that the \himf does not evolve with redshift, so that the local ALFALFA \himf can be assumed to be valid over the redshift range, $0 \leq z \leq 3$, that is considered here. This assumption is motivated by the fact that recent work by \citet{walter_2020_EvolutionBaryons} has found that the cosmic density of hydrogen only evolves by a factor of $\sim 2$ out to $z \simeq 4$, and by the fact that recent work on MIGHTEE early science data has not found evidence for any redshift evolution of the \himf out to $z \simeq 0.084$ \citep{ponomareva_2023_MIGHTEEHFirst}. However, this assumption, as well as the possible impact of an evolving \himf on the results of this paper, is discussed in more detail in Section~\ref{sect: redshift evolution of the HIMF}.

\begin{figure}
    \centering
    \includegraphics[width=\columnwidth]{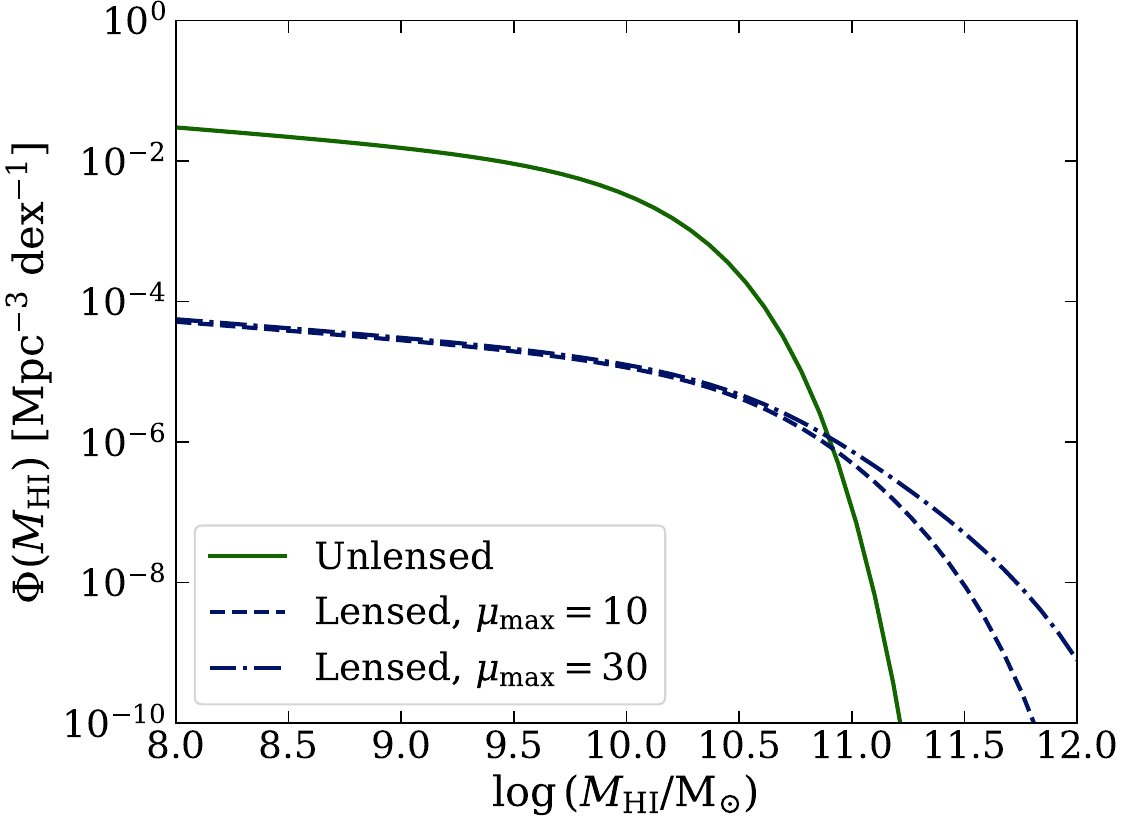}
    \caption[Effect of the magnification bias on the \himf]{The effect of the magnification bias on the \himfe. The green curve shows the unlensed \himf based on the parameters derived from the ALFALFA sample \citep{jones_2018_ALFALFAHI}. The blue curves show the lensed \himf for two values of the maximum magnification (both assume that $\mu_{\text{min}} = 2$). The blue curves are calculated for sources at $z_{\text{S}}=1.5$ an assume a Sheth and Tormen dark matter halo mass function \citep{sheth_2001_EllipsoidalCollapse}.}
    \label{fig:distorted HIMF}
\end{figure}

Figure~\ref{fig:distorted HIMF} shows the distortion of the \himf due to lensing with a toy model. The green curve shows the ALFALFA \himf with the adjusted parameters given above, while the blue curves show the lensed \himf for $\mu_{\text{max}} = 10$ and $30$ calculated according to Equation~\ref{eqn:HIMF distoted by mag bias}. Here, the lensed sources are all assumed to be at redshift $z_{\text{S}} = 1.5$, and $\mu_{\text{min}} = 2$ for both curves. The blue curves intersect the green curve at $\log M_{\text{\mhi}} \sim 11$ indicating that for masses greater than this, the number counts of the lensed population dominates over the unlensed population. It is this point that we aim to utilize in order to find a flux density threshold at which to preferentially select lensed \hi sources.

\begin{figure}
    \centering
    \includegraphics[width=\columnwidth]{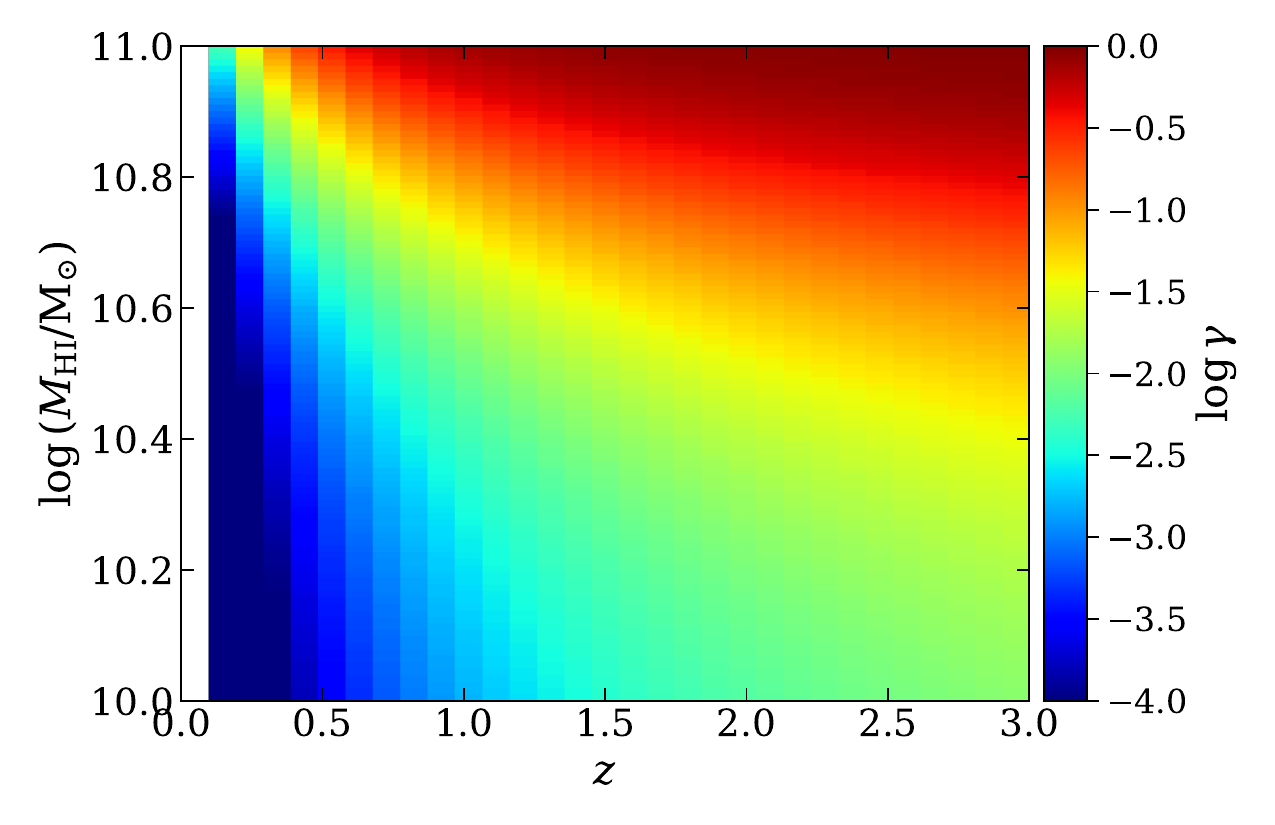}
    \caption[Lens fraction in terms of HI mass and redshift]{The fraction of strongly lensed galaxies (assuming $\mu_{\text{min}} = 2$ and $\mu_{\text{max}}=30$) as a function of $\log$ \hi mass and source redshift. The fraction of galaxies is shown on a $\log$ colour scale.}
    \label{fig:lens fraction}
\end{figure}

Figure~\ref{fig:lens fraction} shows the fraction, $\gamma$, of galaxies that are expected to be strongly lensed as a function of both \hi mass and source redshift. Here, the fraction, $\gamma$ is calculated as the ratio of the distorted \himfe, given by Equation~\ref{eqn:HIMF distoted by mag bias}, to the sum of the distorted \himf and the intrinsic \himf at a given \hi mass and redshift. In calculating the distorted \himfe, it is assumed that $\mu_{\text{min}} = 2$ and $\mu_{\text{max}}$ = 30. The figure shows that it is only at high \hi masses and redshifts that the lensed fraction becomes significant.

%----------------------------------------------------------------------------------------------
\section{Integrated counts of \textsc{Hi} sources}
\label{sect:integrated counts of hi sources}
%----------------------------------------------------------------------------------------------
The integrated source counts of a population can be found by integrating over the volume number density within the relevant limits. Specifically for \hi galaxies the integrated source counts are given by
\begin{equation}
\begin{split}
    N(>S_{\nu}^{\text{peak}}) 
            =    & \int_{0}^{4 \pi} \ud \Omega 
                 \int_{z_{1}}^{z_{2}} \ud z \\
                 & \int_{M_{\text{\mhi}}(S_{\nu}^{\text{peak}}, z)}^{{\infty}} \ud \log{M_{\text{\mhi}}} \Phi(M_{\text{\mhi}}, z) \frac{\ud V_{\text{c}}}{\ud \Omega \ud z},
    \label{eqn:Integrated source counts from HIMF}
\end{split}
\end{equation}
where $M_{\text{\mhi}}(S_{\nu}^{\text{peak}}, z)$ is the \hi mass at a given peak flux density and redshift, $\ud V_{c}/\ud \Omega \ud z$ is the comoving volume element, and $\Phi(M_{\text{\mhi}}, z_{\text{S}})$ is either the intrinsic \himf in the case of the unlensed population, or the distorted \himf in the case of the lensed population. Here and in the rest of the paper the $\log$ notation refers to the base-10 logarithm. Note that $\Phi$ will not have any $z$-dependence in the case of the unlensed population since we are assuming that there is no redshift evolution in the intrinsic \himfe. However, in the case of the lensed population $\Phi'$ does depend on $z$ since the lensing probability depends on the source redshift (see Equation~\ref{eqn:HIMF distoted by mag bias}).

So, in order to calculate the number counts, three quantities need to be specified: first, the maximum magnification for \hi galaxies needs to be estimated, second, the \hi mass needs to be expressed in terms of the peak flux density of the spectral line for the selected channel frequency widths, and third, the redshift interval over which to integrate needs to be chosen. Subsections~\ref{sect: maximum magnification} and \ref{sect: HI mass} discuss the maximum magnification and the \hi mass in terms of the peak flux density. The redshift range is explored in greater detail in Section~\ref{sect: results and discussion} where the results are discussed.

%----------------------------------------------------------------------------------------------
\subsection{Maximum magnification}
\label{sect: maximum magnification}
As mentioned in Section~\ref{sect: distortion himf}, the maximum magnification due to gravitational lensing is limited by the finite size of the source. Hence, the value of the maximum magnification used in Equation~\ref{eqn:HIMF distoted by mag bias} will depend on the angular size of the \hi disc in galaxies, with smaller discs typically having larger magnifications than larger discs, at least for galaxy-galaxy lenses \citep{blecher_2019_FirstDetection, blecher_2024_NeutralHydrogen}.

\citet{perrotta_2002_GravitationalLensing} and \citet{lapi_2012_EffectiveModels} investigate the maximum magnification of extended sources using ray-tracing. \citet{perrotta_2002_GravitationalLensing} consider an ellipsoidal lens potential of a quasi-isothermal sphere, while \citet{lapi_2012_EffectiveModels} consider spherical lens models. They both use ray-tracing to estimate the expected magnification of a source as a function of its position in the source plane. Then, using a de Vaucouleurs brightness profile, they estimate the magnification of the extended source as a function of the distance from its centre, as the ratio of the total lensed brightness to the total unlensed brightness. \citet{perrotta_2002_GravitationalLensing} find that the maximum magnification for sub-mm galaxies with effective radii of 1--10 kpc is in the range 10--30. They note that a value of 10 is fairly conservative and is easily achieved while a value of 30 is only obtained under favourable conditions. \citet{lapi_2012_EffectiveModels} find a slightly higher range of maximum magnifications of 30--50. These values are broadly consistent with modelling of lensed far-IR systems \citep[e.g.][]{vieira_2013_DustyStarburst, negrello_2017_HerschelATLASSample}.

The size of \hi discs, which is typically larger than other baryonic tracers \citep{broeils_1997_Short21cm}, is constrained through the \hi size--mass relation \citep{wang_2016_NewLessons}. This relation constrains the diameter of the \hi discs over five orders of magnitude in \hi mass, giving a range of $\sim 0.3$--300\,kpc for the diameter with a scatter of 0.06\,dex. However, this relation has only been measured in undisturbed systems and only out to redshifts of $z<0.08$ \citep{rajohnson_2022_MIGHTEEHIHI}. At higher redshifts, the size of the \hi discs is uncertain as several factors could affect their extent. One factor is that the size of galaxies is expected to decrease with redshift \citep{gunn_1972_InfallMatter, bouwens_2004_GalaxySize} which would also imply that the diameter of the \hi discs decreases. Another factor is that at earlier cosmological epochs, the effects of turbulence and the ionizing background compete with the effects of minor and major mergers. At these earlier cosmological epochs, the Universe is more turbulent and has a high ionizing background which would tend to decrease the size of the \hi discs, while at the same time, there are more minor and major mergers which would result in more extended \hi discs for a larger fraction of galaxies. Direct mapping of \hi at higher redshifts will help to constrain the $M_{\text{\hi}}$--$D_{\text{\hi}}$ relation. The expectation is that the size would decrease, hence resulting in higher maximum magnifications, however, theoretical work based on cosmological simulations is ongoing.

To estimate the maximum magnification of \hi discs we use the results of \citet{deane_2015_StronglyLensed}. Using $N$-body simulations with ray-tracing, \citet{deane_2015_StronglyLensed} measure the distribution of the magnifications of \hi discs in a simulated area of 150\,$\text{deg}^{2}$ out to a redshift of $z \sim 4$. They found that the majority of sources have magnification factors less than 10, although they did observe magnification factors as high as $\sim 50$ in rare cases. \citet{serjeant_2014_100000} used  a range of 10--30 for the maximum magnification, following the results of \citet{perrotta_2002_GravitationalLensing}. Given all of the above, it seems reasonable to use maximum magnifications in the range 10--30 in this paper, an assumption which will be refined as our knowledge of high redshift \hi improves.

%----------------------------------------------------------------------------------------------

\subsection{An \textsc{Hi} mass--peak flux density model}
\label{sect: HI mass}
The second quantity that is needed in order to calculate the integrated source counts is an expression for the \hi mass in terms of the peak flux density of the spectral line for a given velocity width. Because \hi gas is detected through the hyperfine transition at 21 cm, the number of hydrogen atoms, and hence the mass of the gas, is directly related to the luminosity of the source \citep[see ][]{meyer_2017_TracingHi}. Since the opacity of \hi is typically $\tau \sim 0$, this is true even in the densest environments \citep{braun_1992_PhysicalProperties}. From this, the rest-frame velocity-integrated flux can be related to the \hi mass by
\begin{equation}
    \frac{M_{\text{\hi}}}{\text{M}_{\odot}} \simeq \frac{2.35 \times 10^5}{1+z} \left(\frac{D_{\text{L}}}{\text{Mpc}}\right)^2 \left( \frac{S^{V_{\text{rest}}}}{\text{Jy\,km\,s}^{-1}} \right),
    \label{eqn:HI mass from rest-frame int flux}
\end{equation}
where $D_{\text{L}}$ is the luminosity distance and $S^{V_{\text{rest}}}$ is the rest-frame velocity-integrated flux \citep[e.g.][]{meyer_2017_TracingHi}. However, for our application, we seek a relation between \hi mass and peak flux density, since this work is based on finding a peak flux density selection threshold in a spectral cube at a pre-defined range of channel resolutions, to aid the practical implementation of this approach for spectral lines. This relation can be found through making simplifying assumptions about the shape of the \hi line profile, as well as using the ALFALFA \hi sample \citep{haynes_2018_AreciboLegacy} to relate the velocity width of the \hi line profile to the \hi mass.

\begin{figure}
    \centering
    \includegraphics[width=\columnwidth]{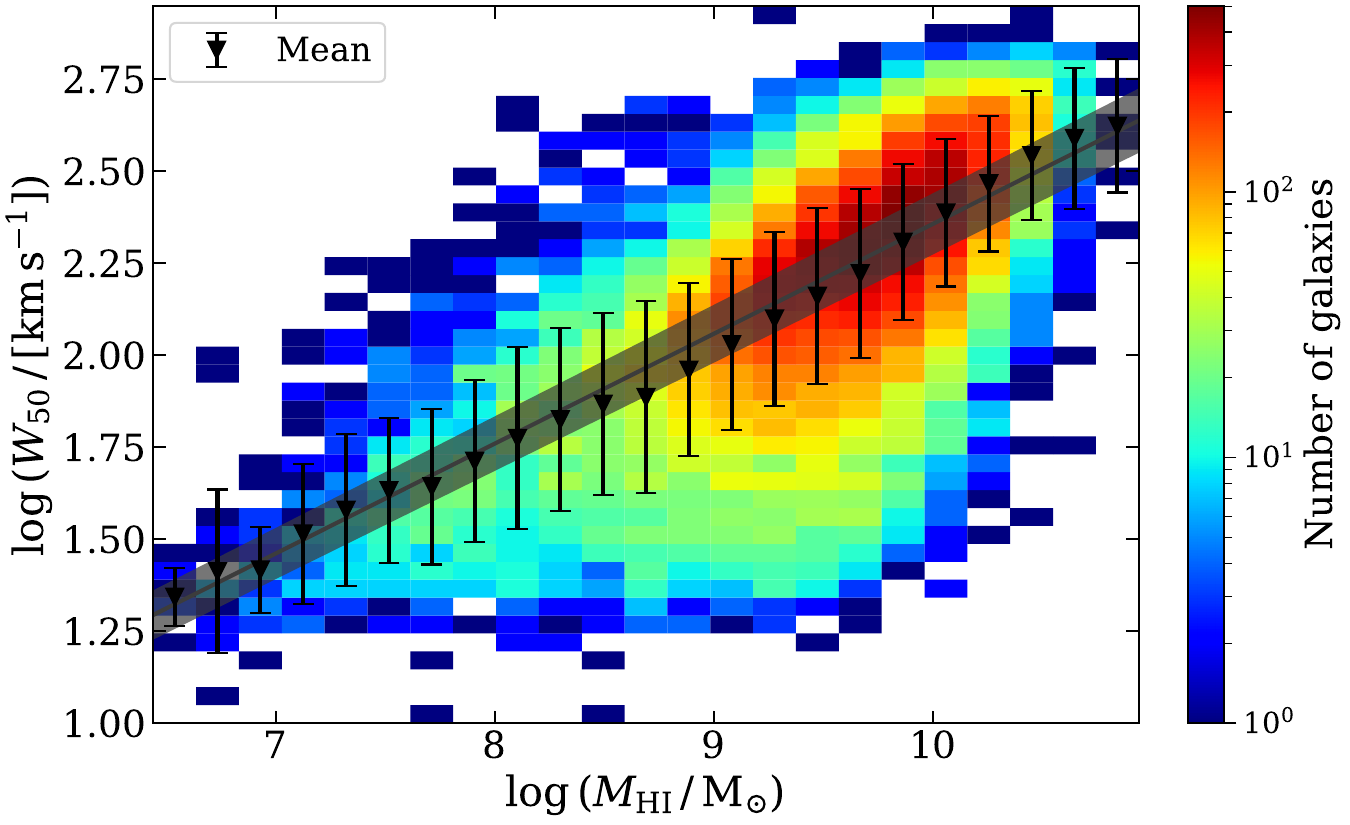}
    \caption[ALFALFA 2D distribution of velocity width and \hi mass]{The 2D distribution of velocity width and \hi mass from the ALFALFA sample for sources with $\log M_{\text{\mhi}}>6.4$. The black triangles indicate the mean $\log W_{50}$ value in each $\log M_{\text{\mhi}}$ bin, while the error bars indicate the standard deviation from the mean in each bin. The grey line indicates a linear fit to the mean values, given by $\log W_{50} = 0.298 \log M_{\text{\mhi}} -0.63$, with the 1$\sigma$ uncertainty on the parameter values indicated by the grey shading.}
    \label{fig:ALFALFA 2D mass-width dist}
\end{figure}

The ALFALFA sample contains $\sim 25\,000$ high quality sources observed in the ALFALFA survey \citep{giovanelli_2005_AreciboLegacy, haynes_2018_AreciboLegacy}. The catalogue provides measures of both the $W_{50}$ velocity width (defined as the velocity width at 50\% of the peak flux density and corrected for instrumental broadening) and the \hi mass for each source in the catalogue. These velocity widths are uncorrected for the inclination of the galaxy. This is desired in our case, since we want to relate the \hi mass to the peak flux density for sources that will also not be corrected for inclination effects. Figure~\ref{fig:ALFALFA 2D mass-width dist} shows the 2D distribution of the ALFALFA sources in bins of $\log W_{\text{50}}$ and $\log M_{\text{\mhi}}$. The triangles indicate the mean $\log W_{\text{50}}$ value in each log mass bin. In order to find a relation between the velocity width and \hi mass, we carry out a linear fit to the mean $\log W_{\text{50}}$ values and the centres of each log mass bin. The best-fitting linear relationship is given by
\begin{equation}
    \log \left( \frac{W_{\text{50}}}{\kmps} \right) = (0.298 \pm 0.005) \log \left( \frac{M_{\text{\mhi}}}{\text{M}_{\odot}} \right) - (0.63 \pm 0.03).
    \label{eqn:logW50--logMHI linear fit}
\end{equation}
This is shown in Figure~\ref{fig:ALFALFA 2D mass-width dist} as the grey curve and the grey shading indicates the $1\sigma$ uncertainty on the parameter values.

The rest-frame velocity-integrated flux can be written explicitly as the integral of the flux density over the rest-frame velocity:
\begin{equation}
    S^{V_{\text{rest}}} = \int S_{\nu} \ud V_{\text{rest}}.
\end{equation}
Due to the rotation and distribution of \hi within the disc, the emission line profile typically has a double peak, if sufficiently inclined and massive. These peaks move close together for intermediate inclinations and lower mass. Here, we make the simplified assumption that the emission line has a boxcar profile. This is the simplest formulation of the shape of the emission line which retains the velocity and total flux of a double horned profile. The assumption is further justified by the fact that we will perform this with a relatively coarse velocity resolution, likely merging the two peaks of any double horn profile. Under this assumption the integrated flux becomes
\begin{equation}
    S^{V_{\text{rest}}} = S_{\nu}^{\text{peak}} W,
    \label{eqn:integrated flux ito peak flux and width}
\end{equation}
where $S_{\nu}^{\text{peak}}$ is the peak flux density of the line and $W$ is the rest-frame velocity width of the line, uncorrected for inclination. Although peak flux density is a brightness and so has units of $\text{Jy\,beam}^{-1}$, in this case the peak flux density is the peak of an integrated spectrum and so is taken to have units of Jy. So then, Equation~\ref{eqn:integrated flux ito peak flux and width} relates this peak flux density to the rest-frame integrated flux for a given rest-frame velocity width.

Finally, combining Equations~\ref{eqn:HI mass from rest-frame int flux} to \ref{eqn:integrated flux ito peak flux and width}, enables us to find a relation between the \hi mass and peak flux density of an \hi source:
\begin{equation}
\begin{split}
    \log M_{\text{HI}} 
            = -1.42 & \left( \log(1+z) - \log(2.35\times 10^5) \right. \\
            & \quad \left. - 2\log(D_{L}) - \log(S_{\nu}^{\text{peak}}) + 0.63 \right).
\end{split}
    \label{eqn:HI mass ito peak flux density}
\end{equation}

\begin{figure}
    \centering
    \includegraphics[width=\columnwidth]{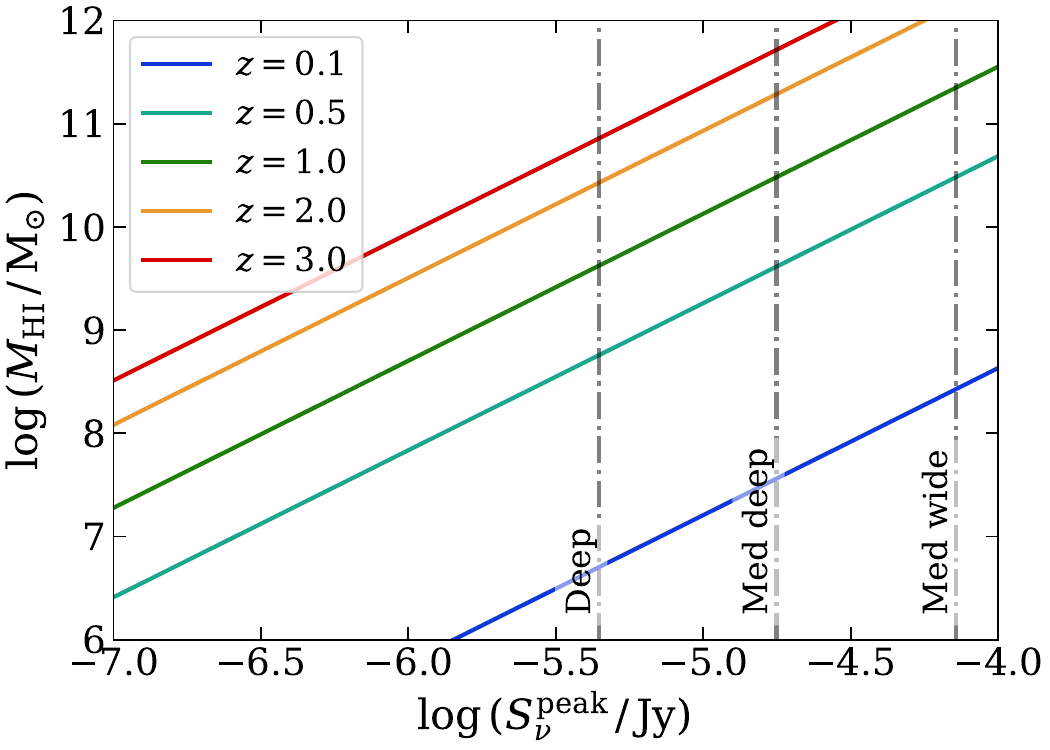}
    \caption[\hi mass as a function of peak flux density]{The derived relation between \hi mass and peak flux density at redshifts $z=0.1, 0.5, 1.0, 2.0$ and $3.0$. The vertical lines indicate the approximate $5\sigma$ sensitivity of the proposed SKA-Mid surveys.}
    \label{fig:HI mass ito peak flux density}
\end{figure}

This relationship is shown in Figure~\ref{fig:HI mass ito peak flux density} at redshifts $z=0.5, 1.0$ and $2.0$.

Note we apply the relation derived in Equation~\ref{eqn:HI mass ito peak flux density} across the redshift range $0 \leq z \leq 3$ even though it is based on the distribution of $W_{50}$ with \hi mass in the ALFALFA sample which is restricted to low redshift sources ($z \leq 0.06$). This is an assumption that will have to be refined in the future as higher-redshift sample sizes increase. Also note that under the assumption of a boxcar profile, the mean flux density of the profile is equivalent to the peak flux density. For other profiles the two will not be equivalent, but Equation~\ref{eqn:HI mass ito peak flux density} will still be valid for the mean flux density. However, we have chosen to keep to the peak flux density description as this is a more intuitive description from an observational perspective.

%----------------------------------------------------------------------------------------------
\section{Results and discussion}
\label{sect: results and discussion}
%----------------------------------------------------------------------------------------------
Section~\ref{sect:integrated counts of hi sources} shows how the general statistical lens selection technique can be applied to the selection of candidate lensed \hi sources. In this section, we discuss the results of the \hi lens selection. The first question to consider is what redshift interval should be integrated over in Equation~\ref{eqn:Integrated source counts from HIMF}. In choosing the redshift interval, both the central redshift and the width of the interval need to be specified. This work is focused on the application to the proposed SKA-Mid \hi surveys described in \citet{staveley-smith_2015_HIScience}. Since the SKA-Mid Band~1 and 2 receivers cover the frequency range that span \hi in the redshift range $0 \leq z \leq 3$, only intervals within in this range are considered here. To explore which redshift intervals this lens selection might yield reasonable surface densities of lensed sources at an accessible flux density, the unlensed and lensed integrated counts are calculated for all the redshift intervals between $z=0$ and $z=3$ for widths of $\Delta z = 0.1$, $0.15$, $0.3$, $0.6$ and $1.0$, resulting in a total of 68 redshift windows for consideration.

\begin{figure*}
    \centering
    \includegraphics[width=\textwidth]{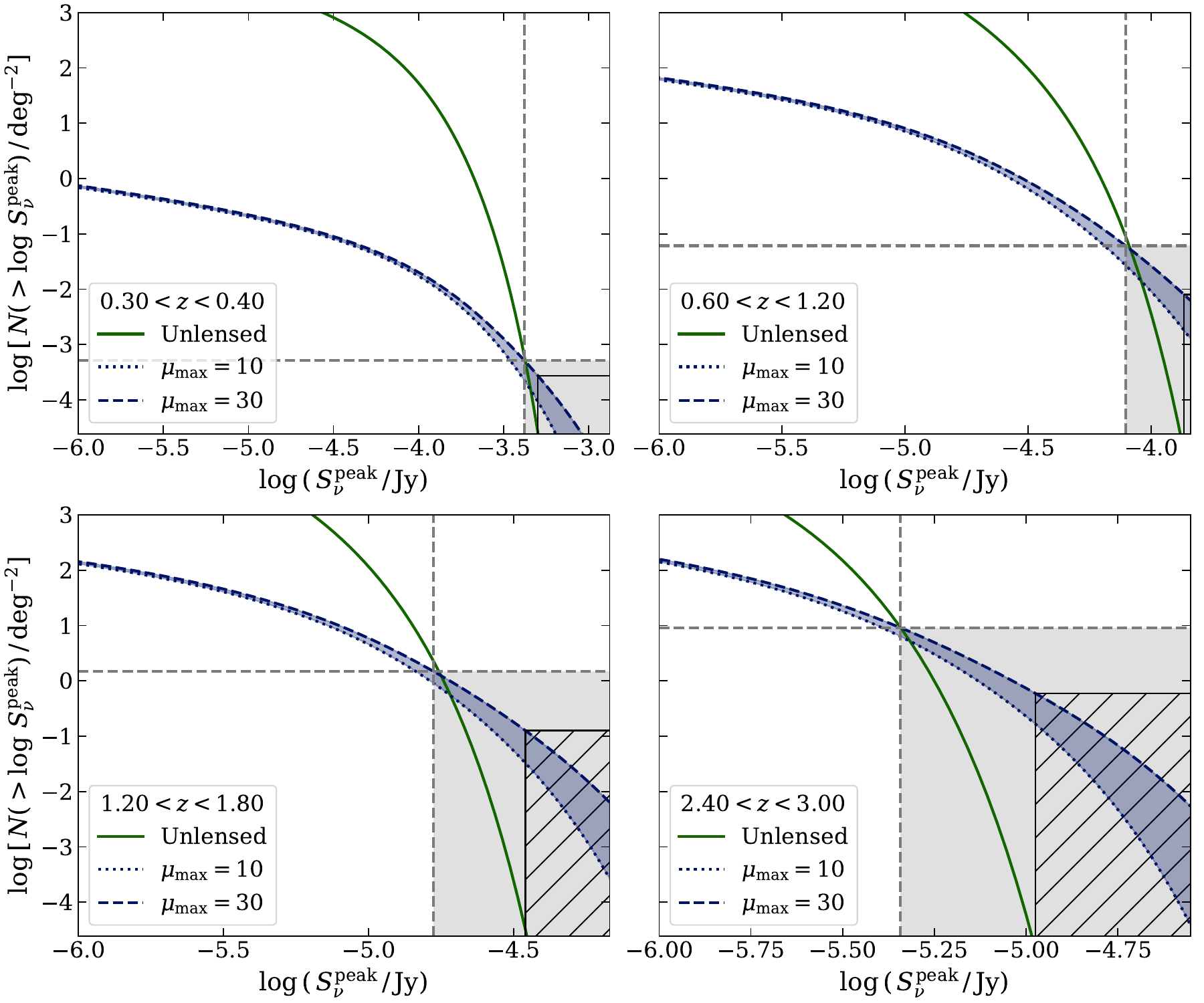}
    \caption[Lensed and unlensed integrated \hi source counts]{Integrated source counts for the lensed (shown in blue) and unlensed (shown in green) populations integrated over different redshift intervals. The lensed source counts are calculated for two values of the maximum magnification, $\mu_{\text{max}} = 10$ and $\mu_{\text{max}} = 30$. The grey shading shows the region where the lensed number counts are equal to or exceed the unlensed number counts, making this the region of efficient lens selection, even if no other multi-wavelength information is used. The hatched region within the shaded region indicates where the unlensed source counts are less than 1 source in the whole sky area. Selecting lenses in this region should have 100\% efficiency.}
    \label{fig:Integrated source counts HI}
\end{figure*}

Figure~\ref{fig:Integrated source counts HI} shows the integrated source counts as a function of the peak flux density for both unlensed and lensed sources, separately, for a subset of the 68 redshift intervals. The integrated source count is the cumulative number of sources that have a peak flux density greater than a given value per square degree on the sky. For the lensed sources, the integrated counts for maximum magnifications of 10 and 30 are shown, in order to give an indication of the range of source counts for different maximum magnifications.

We define the flux density selection criterion to be the flux density value where the lensed integrated source counts start to exceed the unlensed integrated source counts. The region where the lensed source counts exceed the unlensed source counts is indicated by the grey shaded regions in Figure~\ref{fig:Integrated source counts HI}. These plots show that if the redshift interval includes redshifts of $z \lesssim 0.2$, the unlensed counts are always greater than the lensed counts, and there is no flux density selection that will select a sample that contains predominantly lensed sources. This implies that there is a threshold redshift of $z=0.2$ below which this selection technique is not applicable, if no multi-wavelength contaminant removal is applied. A second region of interest is indicated by the hatched region. In this region, the integrated source counts of the unlensed population is less than 1 source in the whole sky for the given redshift interval. This implies that a sample selected at this flux density should consist only of lensed sources.

\begin{figure*}
    \centering
    \includegraphics[width=0.8\textwidth]{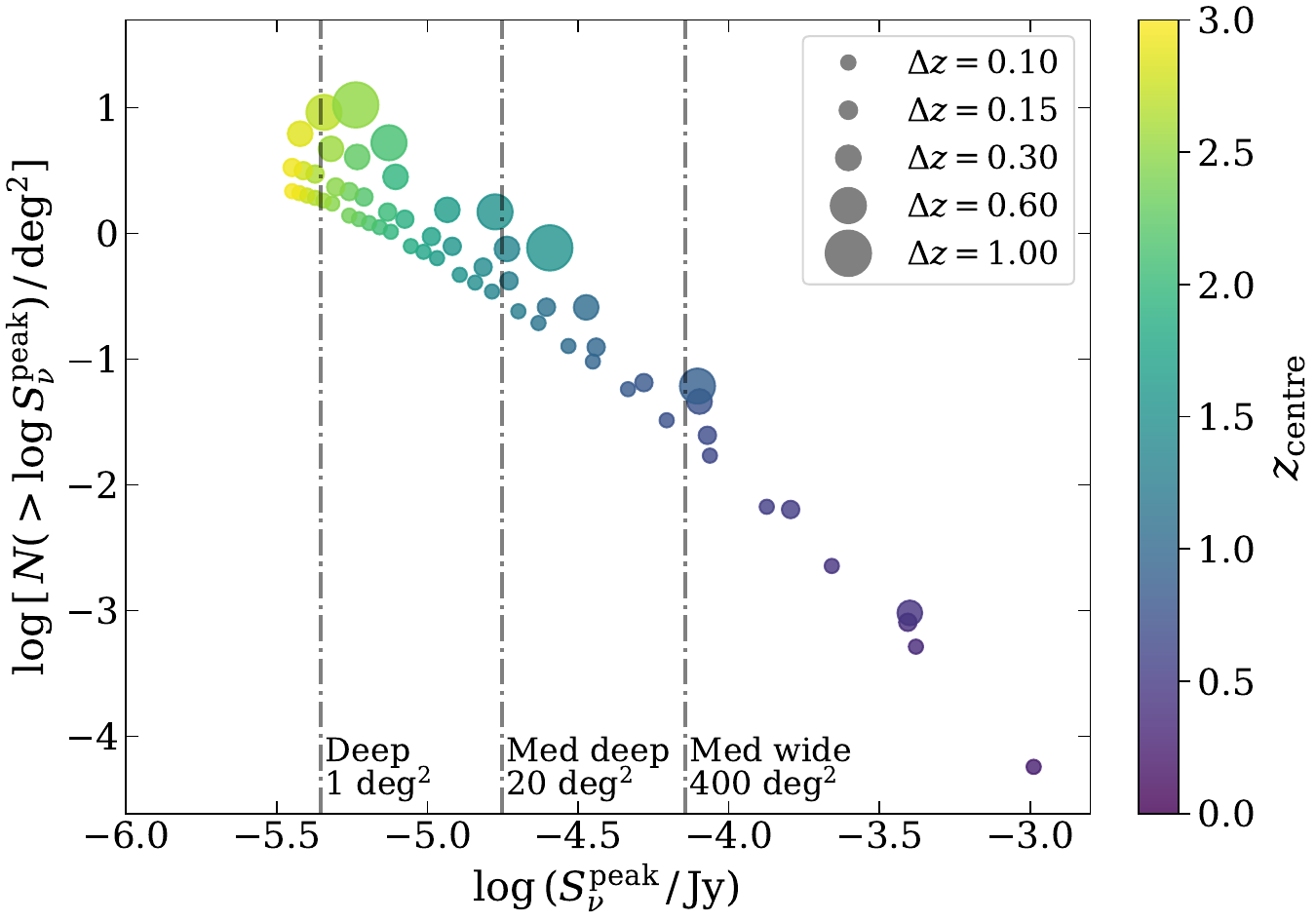}
    \caption[Source count equality points]{Source count equality points for all the redshift intervals considered. The points plotted here are the points in Figure~\ref{fig:Integrated source counts HI} where the grey dashed lines intersect. The centre of the redshift interval is indicated by the colour and the size of the points indicates the width of the redshift interval. The grey dashed vertical lines indicate the approximate 5$\sigma$ sensitivity of the proposed SKA-Mid surveys.}
    \label{fig:Cross-over points HI with surveys}
\end{figure*}

As in \citet{button_2024_EfficientSelection}, we refer to the point where the lensed source counts are equal to the unlensed source counts (shown by the intersection of the dashed grey lines in Figure~\ref{fig:Integrated source counts HI}) as the \textit{source count equality points}. These points are plotted in Figure~\ref{fig:Cross-over points HI with surveys} as a function of the central redshift (shown by the colour bar) and the width of the redshift interval (shown by the size of the markers). As expected, the points move to lower flux densities as the redshift increases, since the received flux from these sources decreases. The points also move to to higher surface densities as the redshift increases, since the probability of lensing increases with redshift. Lastly, the points occur at higher surface densities for larger redshift intervals; this is because a larger volume is included in a larger redshift interval which increases the integrated count. Note that the intervals that include redshifts of $z \leq 0.2$ are not included on the figure since these intervals have no source count equality points.

In addition to the source count equality points, the estimated 5$\sigma$ sensitivities of the proposed SKA-Mid surveys are also shown on Figure~\ref{fig:Cross-over points HI with surveys} by the dot-dashed vertical lines. The estimation of these sensitivities is described in \citet{button_2024_EfficientSelection} and stems from \citet{braun_2019_AnticipatedPerformance} assuming 133 SKA 15-m dishes and 64 13.5-m MeerKAT dishes. This figure enables us to investigate the utility of this approach for selecting lensed \hi sources with the proposed SKA-Mid surveys. As can be seen from the figure, the Medium Wide survey will have the sensitivity to perform this lens selection at redshifts of $z\sim 0.83$. At this peak flux density threshold, the lensed surface density is $\sim 5$ sources in 100\,\sqdegns. Given the proposed area of 400\,\sqdeg for the Medium Wide survey, this selection approach could yield a sample of $\sim 20$ lensed \hi candidates, assuming that the Medium Wide survey will cover the frequency range of the SKA-Mid Band~1 receiver. Similarly, the Medium Deep survey would have the sensitivity to perform the lens selection at redshifts of $z\sim 1.3$, while the lensed surface density at this peak flux density threshold is in the range $\sim 0.2$--$2$ lensed sources per square degree. Since the proposed Medium Wide survey will cover an area of 20\,\sqdegns, this implies that a sample of $\sim 4$--$40$ lens candidates could be selected in this way. Finally, the Deep survey will reach the sensitivity to perform this lens selection out to redshifts of $z \sim 2.6$, while at the peak flux density threshold of the Deep survey, the surface density of lensed sources ranges from $\sim 2$--$9$ sources per square degree. Although the proposed Deep survey would consist of a single pointing covering a field of view of $\sim 1\,\text{\sqdegns}$, at these higher redshifts, the field of view would reach approximately 13.6\,\sqdegns. This implies that it would be possible to obtain a sample of $\sim 27$--$122$ lensed \hi candidates from the Deep survey. These results are summarized in Table~\ref{tab:equality points at surverys}. Although these surveys yield modest lens samples, the lens surface density of the Medium Deep and Deep surveys are higher than that obtained for the \textit{H}-ATLAS survey. For comparison, \citet{negrello_2017_HerschelATLASSample} obtained a sample of 80 candidate lenses over a survey area of $600$\,\sqdegns, giving a lens surface density of 0.13 lens candidates per square degree. Although the sample sizes would be small given the areas of the proposed SKA-Mid surveys, the selection is relatively straightforward with a simple peak flux density cut yielding a sample that should be 50\% lens candidates, prior to any multiwavelength considerations.

\begin{table*}
\centering
    \begin{tabular}{ccccccc}
    \toprule
    Survey & Area & $\log{(5\sigma)}$ & $z$ & $N_{\text{L}}$ (Range) & $N_{\text{L}}$ (Average) & $N_{\text{L}}$ (Total) \\
     & \sqdeg & $\log[S_{\nu}^{\text{peak}}/\text{Jy}]$ & & sources/$\text{deg}^{2}$ & sources/$\text{deg}^{2}$ & \\
    \midrule
    Med Wide & 400 & -4.14 & 0.83 & 0.05 -- 0.06 & 0.05 & 20 -- 24 \\
    Med Deep & 20  & -4.75 & 1.3  & 0.2 -- 2     & 0.6  & 4 -- 40 \\
    Deep     & 1   & -5.35 & 2.6  & 2 -- 9       & 3    & 2 -- 9 \\
    \bottomrule
    \end{tabular}
    \caption[Summary of the equality points at the 5$\sigma$ survey sensitivity levels]{Summary of the source count equality points at the sensitivity levels of the proposed SKA-Mid surveys. The third column provides the 5$\sigma$ sensitivity level of each of the proposed SKA-Mid survey, where the calculation of the sensitivity is described in \citet{button_2024_EfficientSelection}. The fourth column shows the redshift at which the source count equality points reach the flux density threshold given by the survey sensitivity. The final three columns show the range, average and total values of the lensed source counts at these flux density thresholds.}
    \label{tab:equality points at surverys}
\end{table*}

Previous work by \citet{serjeant_2014_100000} also investigated an approach to \hi lens selection with the SKA, based on the effect of the magnification bias on the \himfe. His work considered a survey that was proposed to probe dark energy and that would cover an observing area of 15\,000\,\sqdegns. The estimated sensitivity of the survey assumed the full SKA (i.e. SKA2-Mid) and is based on the work by \citet{abdalla_2010_ForecastsDark}. Using the details provided in both \citet{abdalla_2010_ForecastsDark} and \citet{serjeant_2014_100000}, we estimate the flux density sensitivity used in the predictions by \citet{serjeant_2014_100000} to be $\sim 4\,\mu\text{Jy}$ in a $30\,\kmps$ velocity channel. This is comparable to the sensitivity of the 1\,\sqdeg Deep survey. However, since the work by \citet{serjeant_2014_100000}, the design for the SKA was significantly rebaselined, and it is unclear now whether such a dark energy survey will still be practical with the SKA given the required integration time. The work in this paper, then, provides updated estimates for this lens selection approach based on the currently planned SKA-Mid surveys and incorporating the measured performance of the MeerKAT receivers when estimating the sensitivity of the SKA-Mid surveys.

\citet{deane_2015_StronglyLensed} makes predictions for the number of lensed \hi sources that will be detectable in various SKA and SKA-precursor surveys. They used simulations of a mock observing cone with a field of view of 150\,\sqdeg in order to make their predictions. From this observing cone, they produced a catalogue of simulated \hi sources that are magnified by a factor of 2 or more. Although the assumptions used in the predictions in this work differ from the assumptions used in \citet{deane_2015_StronglyLensed}, by assuming that the flux thresholds in \citet{deane_2015_StronglyLensed} can be converted to peak flux density thresholds using a box-car profile, our predictions can be compared,  and we find that, given the comparatively small cosmological volume over which the ray tracing is carried out, the two sets of predictions agree reasonably well (to within a factor of $\sim 2$--3), given the significantly different approaches, cosmologies, and dark matter properties, amongst other methodological differences.

%----------------------------------------------------------------------------------------------

\subsection{Redshift evolution of the \textsc{HiMF}}
\label{sect: redshift evolution of the HIMF}
This work is based on the local \himf as measured from the ALFALFA survey \citep{jones_2018_ALFALFAHI}. However, the cosmic density of hydrogen has been measured over the range $0 \leq z \leq 4$ using measurements from \hi stacking and DLAs at higher redshifts and has been found to evolve by a factor of $\sim 2$ \citep{walter_2020_EvolutionBaryons}. Since the cosmic \hi density is found from the integral over the \himfe, the normalization and the characteristic mass would also evolve with redshift, if the evolution of the cosmic \hi density is confirmed. There have also been attempts to measure the \himf at $z \sim 0.35$ using the \hi mass--absolute \textit{B}-band magnitude relation and \textit{B}-band luminosity function \citep{bera_2022_HiMass} and at $0 \leq z \leq 0.084$ using direct detections from the MIGHTEE early science data \citep{ponomareva_2023_MIGHTEEHFirst}. The second of these studies, albeit at very low redshifts, did not find any significant evolution in the \himfe, while the first study found that the number of high-mass galaxies ($M_{\text{\mhi}} \gtrsim 10^{10} \, \text{M}_{\odot}$) decreases by a factor of 3.4 from $z=0$ to $z=0.35$, which would indicate evolution in both the characteristic mass and the overall normalization. In contrast to this, recent work by \citet{chowdhury_2024_HIMass}, which is also based on the \hi mass--absolute \textit{B}-band magnitude relation and the \textit{B}-band luminosity function, find that the number density of galaxies with $M_{\text{\mhi}}>10^{10}\,\text{M}_{\odot}$ increases by a factor of 4--5 from $z \simeq 0$ to $z \simeq 1$. 

These recent results indicate that there is mixed evidence for any \himf evolution scenario and that further studies are needed to confirm any evolution or lack thereof. However, if the evolution is confirmed, it will impact the results presented here. Specifically, if the normalization increases, then the surface density of the lensed sources at a given peak flux density threshold will increase. The characteristic mass would impact the peak flux density threshold at which the source count equality points occur for each redshift interval. For a higher characteristic mass, the source count equality points would occur at a higher peak flux densities compared to the non-evolving \himfe. Thus, if the \himf at $z \simeq 1$ is confirmed, the predicted lens source counts would improve if both the normalization and characteristic mass of the \himf increase. A better understanding of the evolution of the \himf at higher redshifts, as well as of the \hi size--mass relation is needed in order to refine these predictions. However, direct detections or lensed detections of high redshift systems are needed in order to further constrain the evolution of these relations before their impact on these results can be investigated. We consider our current estimates as on the conservative side, rooted in well-studied empirical relations and measured SKA precursor performance.

%----------------------------------------------------------------------------------------------
\subsection{Contaminant removal}
\label{sect: contaminant removal}
The selection approach based on the source count equality points is designed to select samples where 50\% of the sources should be lensed \hi candidates and where the remaining 50\% of sources are expected to be unlensed, massive \hi galaxies. Since the goal here is strong lens selection, these unlensed, massive \hi galaxies can be regarded as 'contaminants'. Since truly lensed systems would typically have massive foreground elliptical galaxies in close proximity and/or optical/near infrared arcs, these contaminants could be removed from the sample with additional multiwavelength information. It can be expected that with the up-coming large-area surveys at other wavelengths, such as the Legacy Survey of Space and Time \citep[LSST;][]{ivezic_2019_LSSTScience}, these massive elliptical foreground galaxies would be easily identifiable given their expected mass and the relatively low to intermediate redshifts.

In principle, if the multiwavelength contaminant removal approach is straightforward and effective, samples could be selected so as to contain an even higher ratio of unlensed contaminants to lensed \hi candidates. Although this approach would decrease the flux density threshold of a given sample at a fixed redshift, it would increase the surface density of the lensed candidates, thereby increasing the yield of this statistical lens selection approach.

%----------------------------------------------------------------------------------------------
\subsection{Practical considerations}
\label{sect: practical considerations}
Practically, this lens search would be undertaken by searching the \hi data cubes for pixels that have flux density values that are greater than the given peak flux density threshold. An important consideration would be how to smooth over the spatial and spectral dimensions in order to maximize the signal to noise ratio. The flux density thresholds calculated here assume spatially unresolved galaxies and a velocity width that is related to the \hi mass of the galaxy through the relation between $\log W_{50}$ and $\log M_{\text{\mhi}}$ in the ALFALFA data set, while the sensitivities are calculated for the survey channel sensitivity and a boxcar profile. The signal to noise ratio, and therefore the number of detections, could be further improved with suitably selected smoothing kernels. Additionally, regions near bright continuum sources would have to be masked out if they contain non-Gaussian noise distributions, which would decrease the search volume.

At the SKA-Mid site, radio-frequency interference (RFI) will affect measurements of the \hi 21-cm line between redshifts $0.09 \leq z \leq 0.22$, and to a lesser extent up to $z \sim 0.3$. However, this selection approach will only be compelling at redshifts $z \gtrsim 0.8$, so that the redshift range most affected by RFI in SKA-MID Band 1 and 2 is well below the redshift range that is demonstrated to be compelling in this paper. Moreover, the MeerKAT UHF-band ($\sim 580$--$1000$\,GHz) is a relatively clean band. Hence, RFI is not expected to have a significant impact on the results presented here, although this expectation will have to be confirmed once measurements in the $350$--$580$\,MHz range have been carried out on site.

%----------------------------------------------------------------------------------------------
\section{Conclusion}
\label{sect: conclusion}
%----------------------------------------------------------------------------------------------
As a transitory phase in the baryon cycle, \hi has a significant role in galaxy evolution. Currently, our understanding of the \hi content in galaxies is limited by a lack of direct emission detections at intermediate to high redshifts. Gravitational lensing is a promising tool that could enable direct detections of \hi at intermediate to high redshift within moderate integration times. Additionally, the increase in angular size of spatially resolved lensed \hi sources could enable studies of the \hi kinematics, while lensed \hi systems would provide a different baryonic tracer for selecting dark matter haloes compared to what has been used in the past.

Given the upcoming spectral line surveys proposed for SKA-Mid, which are expected to detect $\sim 10^4$--$10^5$ unlensed \hi sources, this paper investigates a statistical approach to selecting gravitationally lensed \hi sources in survey data. The method used in the paper is based on the distortion of the number counts per \hi mass bin due to the amplification of the \hi signal by gravitational lensing. This distortion is used to find a flux density threshold at which half of the total number of sources are likely to be lensed \hi candidates.

The results indicate that at peak flux density thresholds given by the Medium Wide, Medium Deep and Deep surveys, the average lens surface density is 0.05, 0.6 and 3 lensed sources per square degree, at redshifts of $z=0.83, 1.3$ and $2.6$, respectively. For the proposed survey areas this would yield average sample sizes of 20, 12, and 3 lens candidates for each survey, not taking into account any redshift scaling of the field of view. Although these sample sizes are modest given the proposed areas of the SKA-Mid surveys, the sample selection is simple and should have a 50\% efficiency. Furthermore, including multiwavelength information, including from LSST, will increase the total number of lens candidates dramatically, a topic of future study and optimization. If the evolution of the \himf at $z \simeq 1$ is confirmed as significant, these lens predictions could be even more positive as a higher normalization and characteristic mass would increase both the lensed counts and the flux density thresholds at which the source count equality points occur.

\section*{Acknowledgements}
We thank the anonymous referee for their useful comments on our manuscript. We are grateful to Julie Wardlow, Matt Jarvis, Danail Obreschkow, and Mario Santos for very useful comments and discussions. CBB and RPD acknowledge funding from the South African Radio Astronomy Observatory (SARAO), which is a facility of the National Research Foundation (NRF), an agency of the Department of Science and Innovation (DSI). RPD acknowledges funding by the South African Research Chairs Initiative of the DSI/NRF (Grant ID: 77948). We acknowledge the use of the ilifu cloud computing facility - www.ilifu.ac.za, a partnership between the University of Cape Town, the University of the Western Cape, Stellenbosch University, Sol Plaatje University, the Cape Peninsula University of Technology and the South African Radio Astronomy Observatory. The ilifu facility is supported by contributions from the Inter-University Institute for Data Intensive Astronomy (IDIA - a partnership between the University of Cape Town, the University of Pretoria and the University of the Western Cape), the Computational Biology division at UCT and the Data Intensive Research Initiative of South Africa (DIRISA).

%%%%%%%%%%%%%%%%%%%%%%%%%%%%%%%%%%%%%%%%%%%%%%%%%%
\section*{Data Availability}
No new data were generated or analysed in support of this research.



\bibliographystyle{mnras}
\bibliography{references}

% %%%%%%%%%%%%%%%%%%%%%%%%%%%%%%%%%%%%%%%%%%%%%%%%%%

% Don't change these lines
\bsp	% typesetting comment
\label{lastpage}
\end{document}